\documentclass[twocolumn,showpacs,prb]{revtex4}

\usepackage{graphicx}
\usepackage{dcolumn}
\usepackage{color}
\usepackage{bm}

\begin{document}

\newcommand{\Blue}{\textcolor{blue}}
\newcommand{\Red}{\textcolor{red}}

\preprint{APS/123-QED}

\title{Conductivity engineering of graphene by defect formation\\}

\author{S. H. M. Jafri}
	\altaffiliation{These authors contributed equally to the experimental work}
\affiliation{Department of Engineering Sciences, Box 534, \AA ngstr\"{o}m Laboratory, Uppsala University, 75121 Uppsala, Sweden}
 
\author{K. Carva}
\affiliation{Department of Physics and Materials Science, Box 534, \AA ngstr\"{o}m Laboratory, Uppsala University, 75121 Uppsala, Sweden}
\author{E. Widenkvist}
\affiliation{Department of Materials Chemistry, Box 534, \AA ngstr\"{o}m Laboratory, Uppsala University, 75121 Uppsala, Sweden}
\author{T. Blom}
	\altaffiliation{These authors contributed equally to the experimental work}
\affiliation{Department of Engineering Sciences, Box 534, \AA ngstr\"{o}m Laboratory, Uppsala University, 75121 Uppsala, Sweden}
\altaffiliation{These authors contributed equally to this work}
\author{B. Sanyal}
\affiliation{Department of Physics and Materials Science, Box 534, \AA ngstr\"{o}m Laboratory, Uppsala University, 75121 Uppsala, Sweden}
\author{J. Fransson}
\affiliation{Department of Physics and Materials Science, Box 534, \AA ngstr\"{o}m Laboratory, Uppsala University, 75121 Uppsala, Sweden}
\author{O. Eriksson}
\affiliation{Department of Physics and Materials Science, Box 534, \AA ngstr\"{o}m Laboratory, Uppsala University, 75121 Uppsala, Sweden}
\author{U. Jansson}
\affiliation{Department of Materials Chemistry, Box 534, \AA ngstr\"{o}m Laboratory, Uppsala University, 75121 Uppsala, Sweden}
\author{H. Grennberg}
\affiliation{Department of Biochemistry and Organic Chemistry, Box 576, Uppsala University, 75123 Uppsala, Sweden}
\author{O. Karis}
\affiliation{Department of Physics and Materials Science, Box 534, \AA ngstr\"{o}m Laboratory, Uppsala University, 75121 Uppsala, Sweden}
\author{R. Quinlan}
\affiliation{Department of Applied Science, College of William and Mary, 325 McGlothin Street Hall, Williamsburg, 23187, Virginia, USA}
\author{B. C. Holloway}
\affiliation{Luna Innovations Incorporated, NanoWorks Division, 521 Bridge Street, Danville, 24541, Virginia, USA}
\author{K. Leifer}
\email{Klaus.Leifer@Angstrom.uu.se}
\affiliation{Department of Engineering Sciences, Box 534, \AA ngstr\"{o}m Laboratory, Uppsala University, 75121 Uppsala, Sweden}

\date{\today}

\begin{abstract}
Transport measurements have revealed several exotic electronic
properties of graphene. The possibility to influence the electronic
structure and hence control the conductivity by adsorption or doping
with adatoms is crucial in view of electronics applications. Here,
we show that in contrast to expectation, the conductivity of
graphene increases with increasing concentration of vacancy defects,
by more than one order of magnitude. We obtain a pronounced
enhancement of the conductivity after insertion of defects by both
quantum mechanical transport calculations as well as experimental
studies of carbon nano-sheets. Our finding is attributed to the
defect induced mid-gap states, which create a region exhibiting
metallic behavior around the vacancy defects. The modification of
the conductivity of graphene by the implementation of stable defects
is crucial for the creation of electronic junctions in
graphene-based electronics devices.
\end{abstract}

\maketitle
\section{\label{sec:level1}Introduction\protect\\}
The electronic properties of graphene have been of considerable
interest lately, with several reports of unique structural and
electronic properties.\cite{ageim, bgeim, misha, wehling2007}In the
next phase of research on graphene, it is most likely that we will
witness an increased focus on the tailoring of electronic properties
by means of defects, adatoms and geometrical confinement. The
possibility to open band gaps, induce gap states by defects and the
ability to control the conductivity by chemical and physical means
is hence a crucial step in order to establish graphene as a
competitive material for future electronics. It has in fact already
been shown that a bandgap can be opened in graphene, and that it can
be tuned as a function of carrier concentration.\cite{castro} The
high sensitivity of graphene due to changes in carrier concentration
is related to its low Johnson noise. This property was e.g. used in
the creation of a gas detector capable to detect the adsorption of
single atom. \cite{schedin} Though adsorbed gases sensibly change
the bandstructure of graphene, they tend to be unstable on the
graphene surface. Molecular doping of ${NO_{2}}$ and ${N_{2}O_{2}}$ induce p-type behaviour in graphene. \cite{WehlingNano2007} The maximum
resistance levels are lower for higher concentration of doping. Another road to engineer the bandstructure of
graphene consists in the addition of more stable defects using wet
chemistry methods. This is attractive in view of electronics
applications since such defects are physically more stable than
gases attached to the graphene layer.  In fact, it has been shown
recently that chemically introduced vacancy defects change the
electronic structure of graphene. \cite{coleman}

From the XAS analysis on a graphene layer which was chemically
treated in the same manner as in this work, we have observed a
spectroscopic signature that could be attributed to defect sites in
the graphene planes (i.e. of $\sigma^{\star}$ character). \cite{coleman}
Though the distribution of defects in graphene is
certainly complex and a matter of current studies, these states of
$\sigma$ character related to a defect site could only be oriented
in the graphene plane, and it has been analyzed as being the result
of the presence of a vacancy. Hence we start our study by
considering how vacancy defects influence the electronic structure
of graphene. For this reason we have considered both single vacancy
and di-vacancy in the $\pi$ bonded graphene structure through a
tight-binding model with nearest neighbour interactions. In viewing
the graphene as consisting of two sub-lattices, sub-lattice A and B,
it is known that the effect of an impurity in the A sub-lattice is
manifested in the B sub-lattice. \cite{wehling2007} In particular, a
vacancy in the $\pi$-band, i.e. absence of $\pi$ orbitals at the
impurity site generates mid-gap states for the atoms in the B
sub-lattice located in a neighborhood around the vacancy.
\cite{peres2006} These mid-gap states arise due to symmetry
breaking that removes the equivalent Dirac points in the two
sub-lattices.

\begin{figure}
\includegraphics[width=8.5cm]{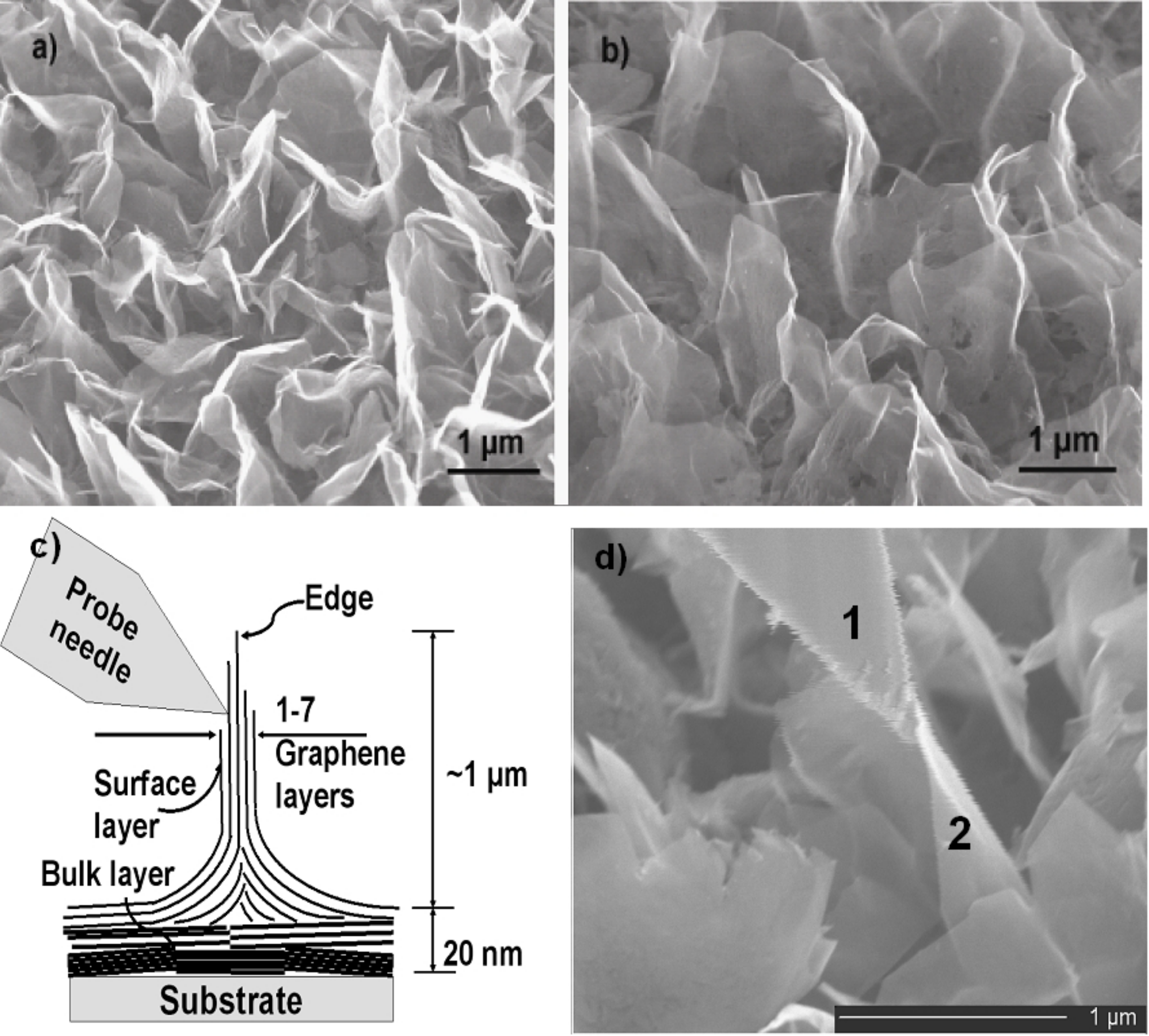}
\caption{\label{fig:epsart} Scanning  electron micrograph of CNS a) on the reference
sample at $52^\circ$ tilt and b) on the acid treated sample at $52^\circ$ tilt, both grown on tungsten substrates 
c) Schematic image of in-situ sharpened probe needle in contact with a CNS 
d)SEM image of the contact between the omniprobe needle(1) and a free standing carbon nano sheet(2).}
\end{figure}

Defects always decrease the mobility of sample compared to its defect-free version, 
while they may increase the number of carriers. The latter effect is especially important here and leads to a strong conductivity rise, 
since for pure graphene there are no carriers at fermi energy. We explore the relevance of defects for controlling
the conducting properties, using both theoretical and experimental
approach. In our theoretical study we investigate the effects on the
local density of states (LDOS) from single vacancies and
di-vacancies by means of real space Green functions (GFs). In order
to understand the influence of the electronic structure on the
conductivity, we calculate the evolution of the LDOS as a function
of distance from the defect position. From there we proceed with
\emph{ab-initio} density functional calculations of the electronic
structure and conductivity. The experimental studies consist of point contact conductivity measurements on free-standing carbon nanosheets (CNS). The CNS consists of few-layer graphene with mono- or double layers at the edges (Fig. ~\ref{fig:epsart}). The samples in this study were grown on a conducting substrate that serves as back contact, greatly facilitating the experimental setup. Defects were introduced by wet chemical treatment, in accordance with previous work. \cite{coleman}

\section{\label{sec:level1} Modelling of defects and prediction of properties \protect\\}

\subsection{\label{sec:level2}Modified local density of states}
The local density of states (LDOS) of a single graphene sheet, both
acid treated and untreated, is modeled in real space by the
dressed graphene Green function (GF)
${\bf G}^r({\bf r},{\bf r}';\omega)={\bf G}_0^r({\bf r}-{\bf r}';\omega)+{\bf G}_0^r({\bf r}-{\bf r}_0;\omega){\cal
T}(\omega){\bf G}_0^r({\bf r}_0-{\bf r};\omega)$; here we discuss the results
for an impurity located at the position ${\bf r}_0=\{{\bf r}_A,{\bf r}_B\}$.
The dressed GF is thus given in terms of the homogeneous GF
${\bf G}_0^r({\bf r}-{\bf r}';\omega)$ with the diagonal and off-diagonal
elements given around the Fermi level $E_F=0$ by
$g_A^r({\bf r}-{\bf r}';\omega)\approx-J_0(k_F|{\bf r}-{\bf r}'|)[2\omega\log(D/|\omega|)+i\pi\omega]/(4\pi\rho
v_F^2)$ and
$g_{AB}^r({\bf r}-{\bf r}';\omega)\approx-J_0(k_F|{\bf r}-{\bf r}'|)(2D+i\pi\omega)/(4\pi\rho
v_F^2)$, respectively, whereas the $T$-matrix is given by
\begin{equation}
{\cal T}(\omega)=[1-\hat{U}{\bf G}_0^r(0,\omega)]^{-1}\hat{U}.
\end{equation}
Here, $\hat{U}=(U_A, U_B) $ is the impurity potential matrix,
where $U_A\ (U_B)$ is the scattering potentials at ${\bf r}_A\
({\bf r}_B)$, whereas $J_0(x)$ is a Bessel function of the first kind,
$v_F$ is the Fermi speed, $k_F$ is the Fermi wave vector, and
$\rho$ is the graphene planar density and is related to the energy
cut-off $D\sim5$ | 10 eV through $4\pi\rho v_F^2=D^2$. Although this
approximation does not describe the modified LDOS quantitatively
correct, it gives a sufficient qualitative account.
First we consider a single impurity located at ${\bf r}_A$. As we are
considering a vacancy, we let $U_A\rightarrow\infty$. The
perturbation to the LDOS in the A sub-lattice then becomes $\delta
N_A({\bf r},\omega)=-\Im \delta G_A^r({\bf r},{\bf r};\omega)/\pi\approx\omega
J_0^2(k_F|{\bf r}-{\bf r}_A|)/(4\pi\rho v_F^2)$, showing that the Dirac
points remain in the A sub-lattice. The perturbed LDOS in the B
sub-lattice, on the other hand, becomes
\begin{eqnarray}
\delta N_B({\bf r},\omega)\approx
    \frac{J_0^2(k_F|{\bf r}-{\bf r}_A|)}{4\pi\rho v_F^2} \Im \frac{{(2D+i\pi)^2}}{\omega[2\log(D/|\omega|)+i\pi]},
\end{eqnarray}
which is clearly diverging at the Fermi level. This divergence
signifies the existence of mid-gap states for the carbon atoms in the B
sub-lattice. Using more sophisticated approximations, which removes
the unphysical singularity, show that the mid-gap states are indeed
present in the B sub-lattice. \cite{peres2006,wehling2007}
\subsection{\label{sec:level2}Calculations of electronic structure and resistivity}
We performed \emph{ab-initio} density functional calculations of the
graphene electronic structure in presence of single vacancies and
di-vacancies. The calculations employed the projector augmented wave
method, and the plane wave cut-off energy was set to 750 eV. The
generalized gradient approximation (GGA) was used for the treatment
of exchange-correlation functional and a lateral 8x8x1 supercell of graphene with a divacancy was considered for the calculations. 
This yields an effective vacancy concentration of 0.0156. The geometry of
the supercell in presence of the defect was optimized using
Hellmann-Feynman forces with a tolerance of 0.01 eV/ \AA. A
$\Gamma$-centered 3x3x1 set of k-points was used in the
Monkhorst-Pack scheme with a Gaussian broadening of 0.2 eV.

In the calculations of the graphene conductivity we used an
all-electron scalar-relativistic version of the tight-binding linear
muffin-tin orbital (TB-LMTO) method \cite{r_97_tdk} within the local
spin-density approximation \cite{r_72_vbh} (LSDA) to the density
functional theory. \cite{r_64_hk} The valence basis consisted of
$s$-, $p$-, and $d$-orbitals. The atomic sphere approximation (ASA)
was employed here, which required an additional empty sphere to be
present in the graphene layer per two carbon atoms and other empty
spheres between graphene layers with a spatial distribution similar
to the one already used successfully for graphite.\cite{r_96_Solan_GrASA_14ES_DOS} The integrations over the 2D BZ
were performed on a uniform mesh of about 5000 ${\bf
k}_{\|}$ points. Such high density of ${\bf k}_{\|}$-mesh is crucial
for the accurate description of states near the Dirac point. 

The random distribution of defects in the system is described by means
of the coherent potential approximation.
(CPA)\cite{r_67_ps,r_68_vke} This single-site approximation has
been shown to be successful in predicting properties of systems with
random disorder, thus averaged over all possible combinations of
impurity distributions. The CPA can deal with systems with arbitrary
low concentration of impurities while its numerical requirements
remain modest compared to the supercell method for disordered
systems. The CPA has been used in other studies of graphene, 
where it was argued that CPA gives very good results for the description of the physical properties of graphene \cite{ r_06_Nilss_DisGrene_CPA, r_05_Peres_disGrene_CPA_FM}. 
The question concerning applicability of the CPA near Dirac point was studied by Skrypnyk, Loktev \cite{r_07_Skryp_ImpurDiracPSmearing}. 
They have noted that only renormalized approaches as CPA can be efficient close to the van Hove singularities in the spectrum. 
Furthermore, it was found that in the vicinity of the Dirac point a correction to the CPA self-energy is needed. 
We study DOS and resistance only for systems with a finite concentration of impurities, which naturally induces a shift of the Fermi level away from the Dirac point. 
Hence, CPA is applicable for the present study. \cite{r_06_ctkb_vertex} The conductance was calculated from
the Kubo linear response theory.

\section{\label{sec:level1}Experimental verification\protect\\}
\subsection{\label{sec:level2}Graphene model system}
The results from theory are predictions for an idealized material which is not in contact with any surroundings. In order to experimentally verify the predictions, we have used well-characterized carbon nanosheets ( Figs.~\ref{fig:epsart}(a) and ~\ref{fig:epsart}(b)). These CNS were grown by plasma-assisted chemical vapor deposition (CVD) on a conducting substrate as a model system for graphene.\cite{Wang} The CNS, consist of 1$\mu$m graphene sheets protruding from a 20 nm layer of graphite. The estimated thickness is 2-3 graphene layers at the top and about 7 monolayers in the vicinity of the substrate. The thus supported CNS is allowing both wet chemical treatment\cite{coleman} and point contact measurements without the adverse effects associated with removal of the CNS from the growth substrate and the deposition and contacting required for the conductivity study (schematics is shown in Fig.~\ref{fig:epsart}(c)). \cite{French} Compared to graphene adhered and contacted i.e. on top of SiO2, this set-up is likely to reduce the influence of substrate charges and impurities from graphene processing which might shade the effect of defects on the conductivity.
\subsection{\label{sec:level2}Sample preparation}
The defects were introduced in the CNS by a treatment with concentrated aqueous HCl as described,\cite{coleman} the CNS were exposed to 35\% concentrated aqueous HCl at $90^\circ$C for 3 hours. The HCl was removed from the flasks using a pipette and the samples were treated with deionized water for 10 h at $90^\circ$C, followed by a thorough rinse with RT deionized water and drying in air at $150^\circ$C for 3 min. In order to assess the effect of this acid treatment, a reference sample was prepared using deionized water instead of the aqueous HCl then followed by the same wash - rinse - dry protocol. Water, in contrast to aqueous HCl, cannot induce permanent defects. The samples are in the following referred to as "acid-treated" and "reference", respectively. All samples, includning the as-deposited, were handled in air.

\subsection{\label{sec:level2}DC measurement setup}
The CNS samples were electrically contacted with an
Omniprobe nano-manipulator inside a FEI Strata DB235 focused ion
beam/scanning electron microscope (FIB/SEM). The Current-Voltage (I-V) measurements of the CNS were performed by using
a Keithley 6430 source meter. The tungsten tip of the
manipulator is in-situ ion polished in the FIB providing a better
definition of the tip/graphene contact area as well as a reduction
of the oxygen on the W surface. 

In our nano-manipulation experiment, the CNS are contacted with a sharp tip
at room temperature. The probe needle is manoeuvred in the
x, y and z-directions until it contacts a CNS. All movements are
carefully monitored in real time by continuously scanning the
electron beam in the FIB/SEM. The tip/CNS contact in Fig.~\ref{fig:epsart}(d) is
established in the upper part of the CNS (for schematic
representation of tip and carbon nano sheets, see Fig.~\ref{fig:epsart}(c)). 
The tip/CNS contact can be precisely observed through a slight bending of the CNS in the realtime scanning electron
micrograph (SEM) images. The voltage was scanned from -1 to 1V
in 30-40 seconds with no applied gate voltage. A reference SEM image
was recorded before and after each measurement. When the tip/CNS contact is not
sufficiently strong, occasional drift and tip vibrations led to a (I-V)
temporary detachment of the tip seen as a current drop in the 
curves. Such I-V curves were not considered in the further
evaluation. In this measurement geometry CNS are available
in abundance with a surface density of $10^8$ cm$^{-2}$. Therefore,
the conductivity of a large quantity ($>$ 400) of CNS could be
assessed and consequently, the data can be extracted with a high
statistical precision. Three to five measurements are taken at the same location. The magnetic field
inside the FIB/SEM chamber is found to be approximately 66mT in Ultra High Resolution Mode. 

\section{\label{sec:level1}Result and disscussion\protect\\}
\subsection{\label{sec:level2}Influence of Vacancy defect on Density of States}
\begin{figure}
\includegraphics[width=8.5cm]{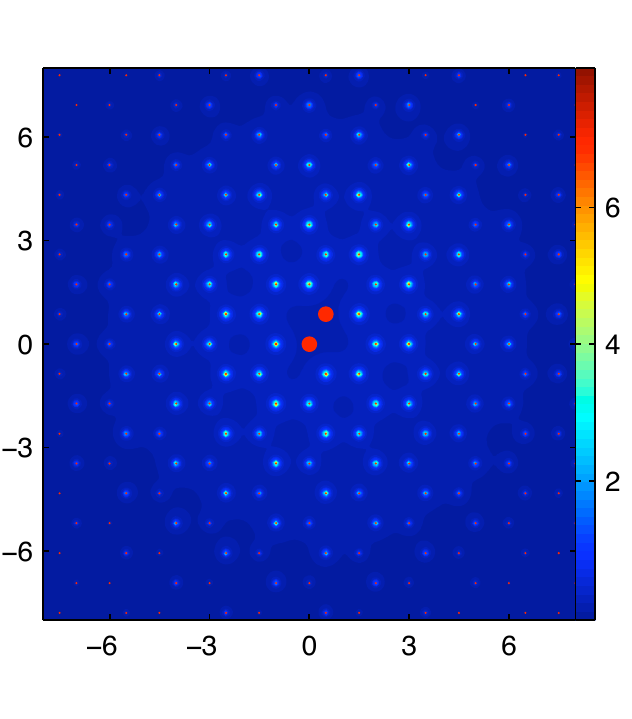}
\caption{\label{fig:epsart1} (Color online) Real space distribution of the correction
$\delta N({\bf r},E_F)=-\Im {\sum_{i=A,B}}$ $\delta G_i({\bf r};E_F)/\pi$
to the LDOS (arb. units) at the Fermi level in presence of the
di-vacancy located at ${\bf r}_A=0$ and ${\bf r}_B=a(1,\sqrt{3})/2$ (large
red dots). The coordinates are given in units of the lattice
parameter $a$.}
\end{figure}

\begin{figure}
\includegraphics[width=8.5cm]{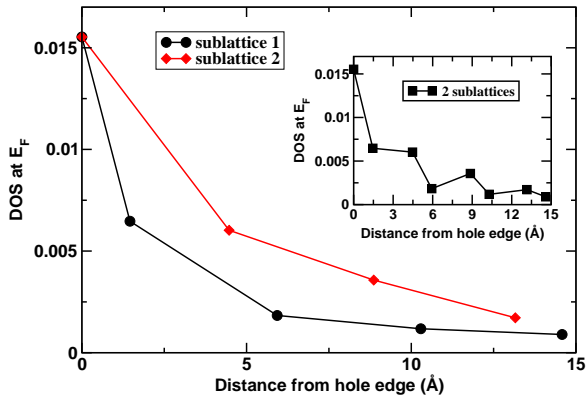}
\caption{\label{fig:epsart2} {Calculated Density of states (DOS) at the Fermi level
for C atoms, as a function of distance from a
di-vacancy of graphene. The values are shown for both sublattices. In the inset we show the same data, for both sublattices in one single graph.}}
\end{figure}

In our first approach to understand the electronic structure in
functionalized graphene, the LDOS of a single graphene sheet, both
defect free and defect containing, is modeled by means of real space
GFs. Consider a single impurity
located at $\bf r_A$ in sub-lattice A. Although the LDOS is slightly
modified in sub-lattice A, the Dirac point remains. On the other
hand, in sub-lattice B, the effect from the vacancy is more
pronounced in that mid-gap states arise from the carbon atoms.
\cite{peres2006, wehling2007} The emerging mid-gap state suggests
that the B sub-lattice becomes metallic in presence of the vacancy
in the A sub-lattice. Generalizing the above consideration to the
case of a di-vacancy, with two adjacent vacancies of which one is in
each sub-lattice, shows the emergence of the mid-gap states in both
sub-lattices for atoms in a neighborhood around the di-vacancy. This
conclusion indicates that graphene becomes metallic in a
neighborhood of the di-vacancy. The length scale associated with the
metallicity around the di-vacancy is of the order of a few lattice
parameters, Fig.~\ref{fig:epsart1}, which is due to the spatial power law decay
of the influence from the impurity. 
In order to substantiate the model considerations above we have
performed density functional calculations using the Vienna Ab-initio
Simulation Package for a graphene sheet with a di-vacancy.
\cite{vasp1,vasp2} The value of the DOS at the Fermi level
($E_{F}$), which signifies the existence of a metallic conducting
component, is shown in the Fig.~\ref{fig:epsart2}. A relatively large peak
arising from the $p_{z}$ orbital develops at the Fermi level for an edge
atom around the di-vacancy. If one moves away from the defect
site, the DOS at $E_{F}$ decreases, but the metallic component
extends over several lattice sites around the defect (more than 15 {\AA} away from the defect site). In the inset of Fig. 3, we show the DOS at the Fermi level for both sublattices, suggesting an oscillating behaviour. A closer inspection (main body of Fig. 3) reveals that this is simply due to that the two different sublattices in the graphene sheet show different decay rates away from the defect center. Our
ab-initio calculations are in agreement with the model calculations
discussed above, that is, vacancies introduce a metallic component
in the electronic structure, and this metallicity,
extending to several lattice constants, should be sufficient to
enhance the conductivity of the graphene.
\subsection{\label{sec:level2}Influence of Vacancy defect on resistivity}
The central aspect of this paper is the conductance of graphene as a
function of vacancy defects, which we describe in the following ab-initio
calculations. For simplicity we only considered
single C vacancies in these calculations, but in light of the
discussion above, we expect the results to hold also for other
vacancy geometries.

\begin{figure}
\includegraphics[width=8.5cm]{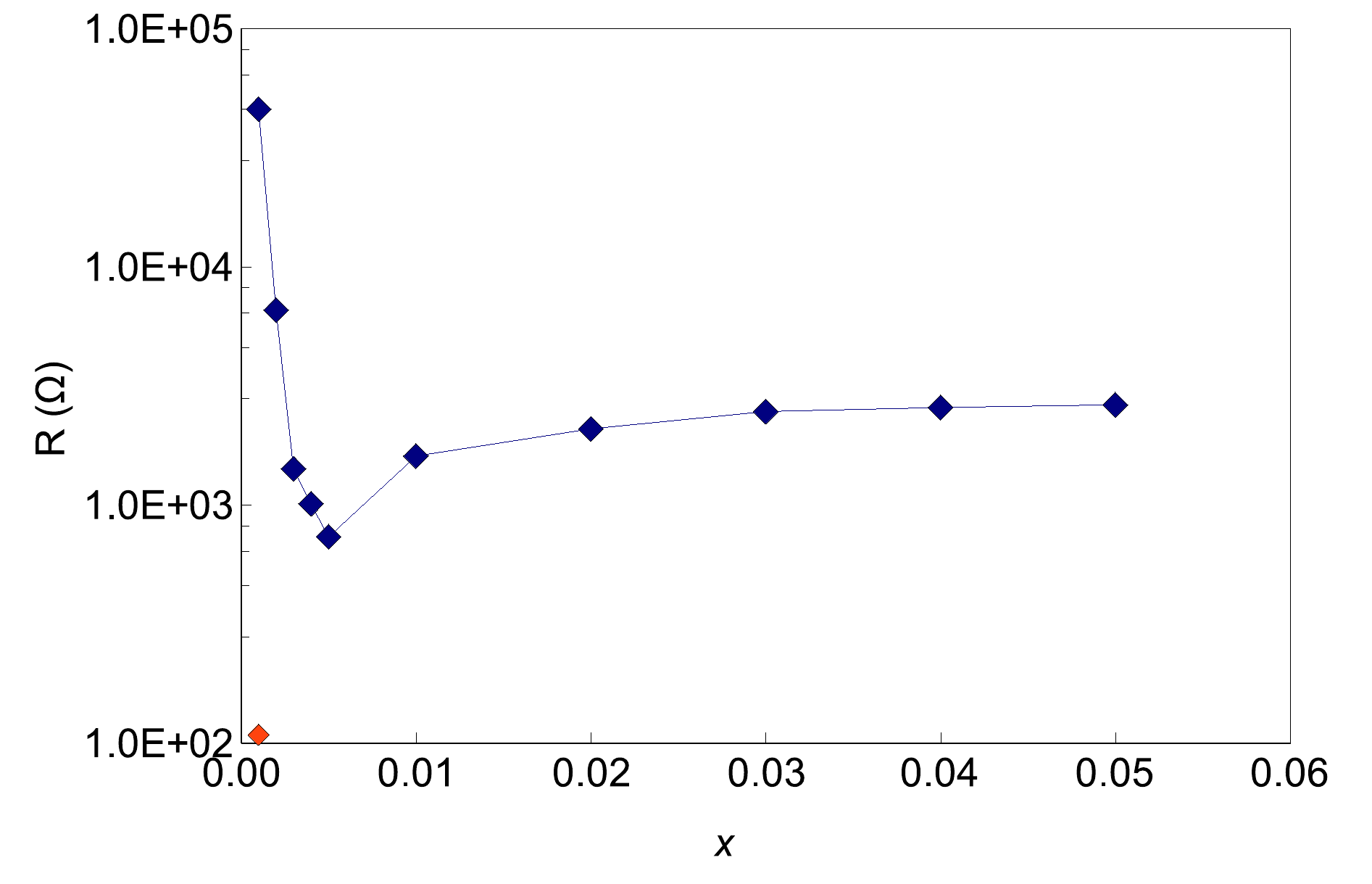}
\caption{\label{fig:epsart3} Calculated resistance of a single layer of graphene as a
function of defect concentration. The red diamond represents a calculated resistance from a model consideration with a low amount of defects but a relatively large shift of the Fermi level (see text).}
\end{figure}

The results of our calculations are shown in Fig.~\ref{fig:epsart3}. It is obvious
that as the concentration of vacancies grows, the 0K temperature
resistivity decreases with at least one order of magnitude compared
to the material containing no vacancies. In the concentration regime of x $\le$ 0.005 it is seen from
Fig.~\ref{fig:epsart3} that the conductivity depends exponentially with respect to defect concentration. This is actually a normal result for regular semiconductors \cite{r_73_Seeger_semicon}, and can be explained from a Fermi-Dirac distribution of electron states. For a larger concentration
of vacancies ($x > 0.005$) it is found that the resistivity slightly
increases (Fig.~\ref{fig:epsart3}), and appears to saturate close to a constant
level for a defect concentration of 3-5 \%. These results verify our
conclusions drawn from the character of the electronic structure in
presence of vacancies.
The results in Fig.~\ref{fig:epsart3} should therefore be seen as a transition from
a semi-metallic regime (or rather a regime where the effects of
Zitterbewegung dominate) with limited conductivity,\cite{bgeim, kats}
to a regime of highly conducting graphene. This
transition is driven by defects via the creation of mid-gap states,
which produce an extended regime of metallic character. Once this
regime has been fully established, the addition of more defects
produce scattering centers which reduce the conductivity, resulting
in a more conventional behavior. The connection between defects and
resistivity in graphene is very intricate. Similar changes in the
density of states were observed in graphene nanoribbons, where
states in the midgap appear at interfaces.\cite{Xu}

Defects cause not only impurity states, but also accept or donate charge, thus causing a shift of the Fermi energy in the studied system. We have compared the Fermi levels for different impurity concentrations in order to quantify the influence of the Fermi level shift on conductivity. In order to distinguish between the contributions from the Fermi level shift and impurity scattering on the conductivity, we have considered a model system as follows. The Green function of the model system is identical to that of a graphene with a very low concentration of defects (0.1 \%), but the Fermi level corresponding to a much larger concentration of defects (metallic regime, i.e. 1 \% defect) was used in the spirit of the rigid band approximation. The system will further be denoted as graphene II. Therefore, the difference between the graphene II system and the original graphene consists only in a change of the Fermi energy. The calculated resistance of graphene II is shown as a (red) diamond in Fig.~\ref{fig:epsart3}. Clearly the conductivity of graphene II is greatly enhanced compared to the original graphene with an impurity concentration of 0.1\%, although their only difference is 0.016 Ry change of the Fermi energy.

In a metallic regime the residual resistance originates from impurity scattering and it is thus expected to be proportional to the impurity concentration. Graphene with $x=1\%$ and the graphene II with $x=0.1\%$ (Fig.~\ref{fig:epsart3}) should differ mainly by the amount of impurity scattering as compared to the different conductivity regime of graphene with $x=0.1\%$. When the ratio of the resistances between the two systems with the same Fermi energy and different impurity concentration i.e. graphene at $x=1\%$ and the graphene II at $x=0.1\%$ is evaluated, it indeed approximates the expected proportionality for metals. Therefore the intial decrease of the resistivity curve in Fig.4 is due to an increased metalic component of the electronic structure, with negligible influence from impurity scattering. At higher defect concentrations (0.5 \% and above) the effect of the impurity scattering becomes important, and the balance between these two effects causes the resistivity to grow with large defect concentrations.

In the experimental situation, the dangling bonds around the defects
in graphene are most likely to be saturated with hydrogen, oxygen or
hydroxide groups. In order to account for this, we have carried out
density functional theory calculations saturating the dangling bonds
around di-vacancies with H and/or O atoms. The LDOS of the carbon
atoms around the defect at the Fermi level reaches, as in the
calculations of vacancies with dangling bonds, finite values
suggesting that metallicity persists upon addition of H and/or O to
the carbon atoms in the neighborhood of the vacancies.
\subsection{\label{sec:level2}IV characterization of Carbon nanosheets}
The presence of oxygen and absence of chlorine was verified by x-ray
photoelectron spectroscopy in model CNS system. In our XAS measurements, we observe
in-plane, $\sigma ^*$ -type orbitals appearing after the acid
treatment which indicates the presence of vacancies. In fact, when
graphene is damaged for example by electron irradiation, both,
single and di-vacancies were found.\cite{Hashimoto, Meyer} But as
shown above, the appearance of midgap states is valid for both
single and di-vacancies and therefore, at this stage, our transport
experiment is only sensitive to the presence of defects, but not to
their detailed character.
\begin{figure}
\includegraphics[width=8.5cm]{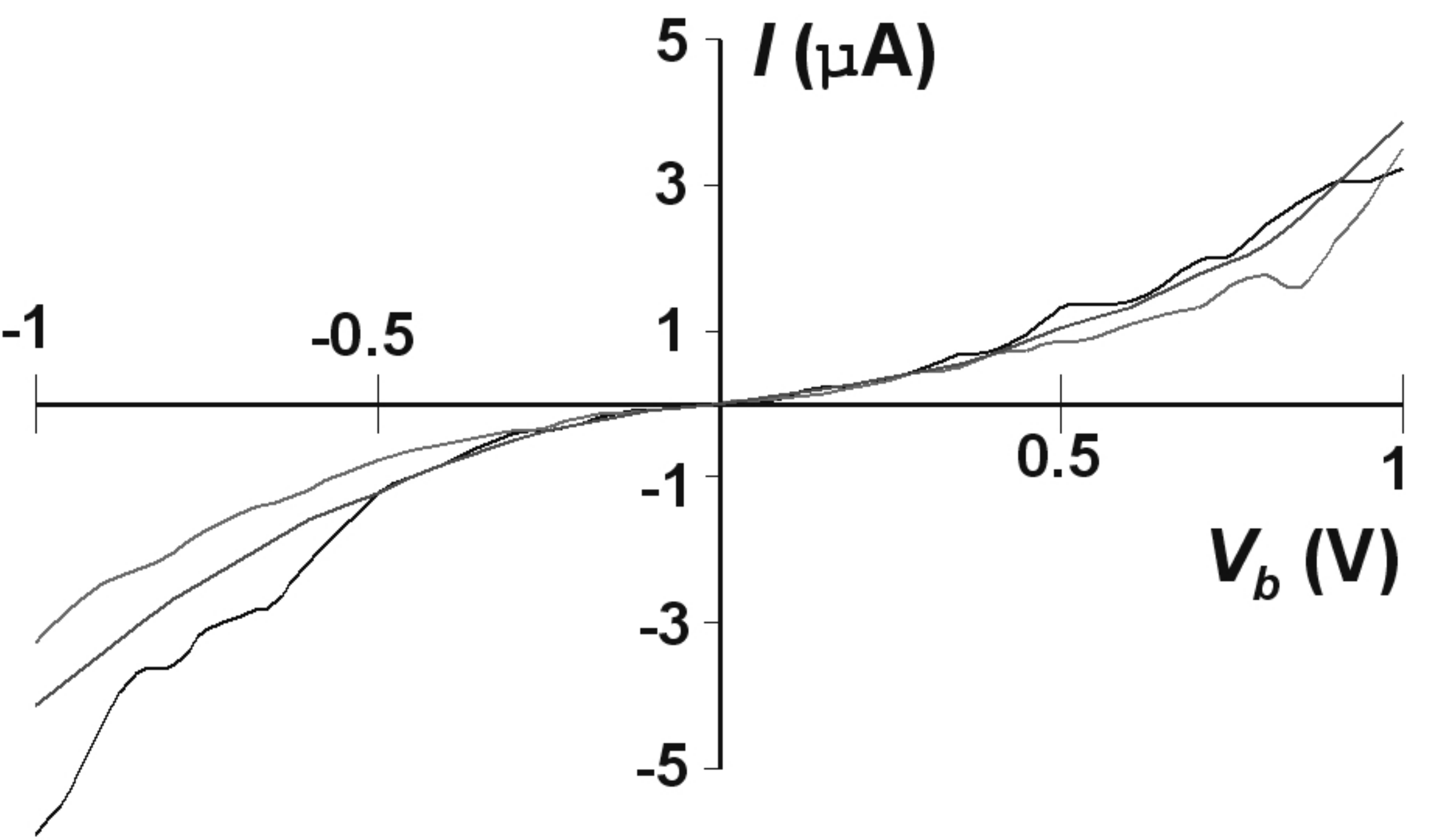}
\caption{\label{fig:epsart4} I-V characteristics
in forward and reverse bias of the water treated (reference) sample.}
\end{figure}
Fig.~\ref{fig:epsart4} shows three I-V measurements performed on the reference sample and Fig.~\ref{fig:epsart5} shows three I-V curves of  acid treated (defect induced) sample, demonstrates that the I-V curves on individual CNS are
reproducible. The I-V curves of Fig.~\ref{fig:epsart4} and Fig.~\ref{fig:epsart5} are symmetric and
non-linear, where the reference sample shows currents in the range
of 0.1-10 $\mu$A  and the acid treated sample in the range of
0.01-1 mA at 1 V. The currents in the
as-deposited sample are about 2 times lower than in the reference
sample.  In order to evaluate the curves quantitatively, we have
measured the resistance at bias voltages from -0.6V to 0.6V. Every
recorded I-V curve results in a number of resistance values taken at
different bias voltages V$_B$. Both negative and positive voltages
are considered for the evaluation since no systematic asymmetry was
seen in the I-V curves for opposite bias. The resistances are
plotted in histograms for each voltage value (Fig.~\ref{fig:epsart6}). The scatter of the resistance data between the
different CNS on the same sample are due to changes in size of the
CNS and of the contact point of the needle.

The measured data collected in resistance histograms were
fitted with a log-normal distribution, which is used to determine
the average resistance as well as the standard error of the
resistance at each voltage. These resistance values and their
associated errors are shown in Fig.~\ref{fig:epsart7}, as a function of bias voltage.
The average resistance of the reference sample is about 50 times
higher than the resistance of the acid treated sample. This
change in resistance is much larger than the standard error of our
measurement (Fig.~\ref{fig:epsart7}), and we can safely conclude that our
described method of chemical treatment can be used to tune the
conductance of CNS with more than one order of
magnitude.
\begin{figure}
\includegraphics[width=8.5cm]{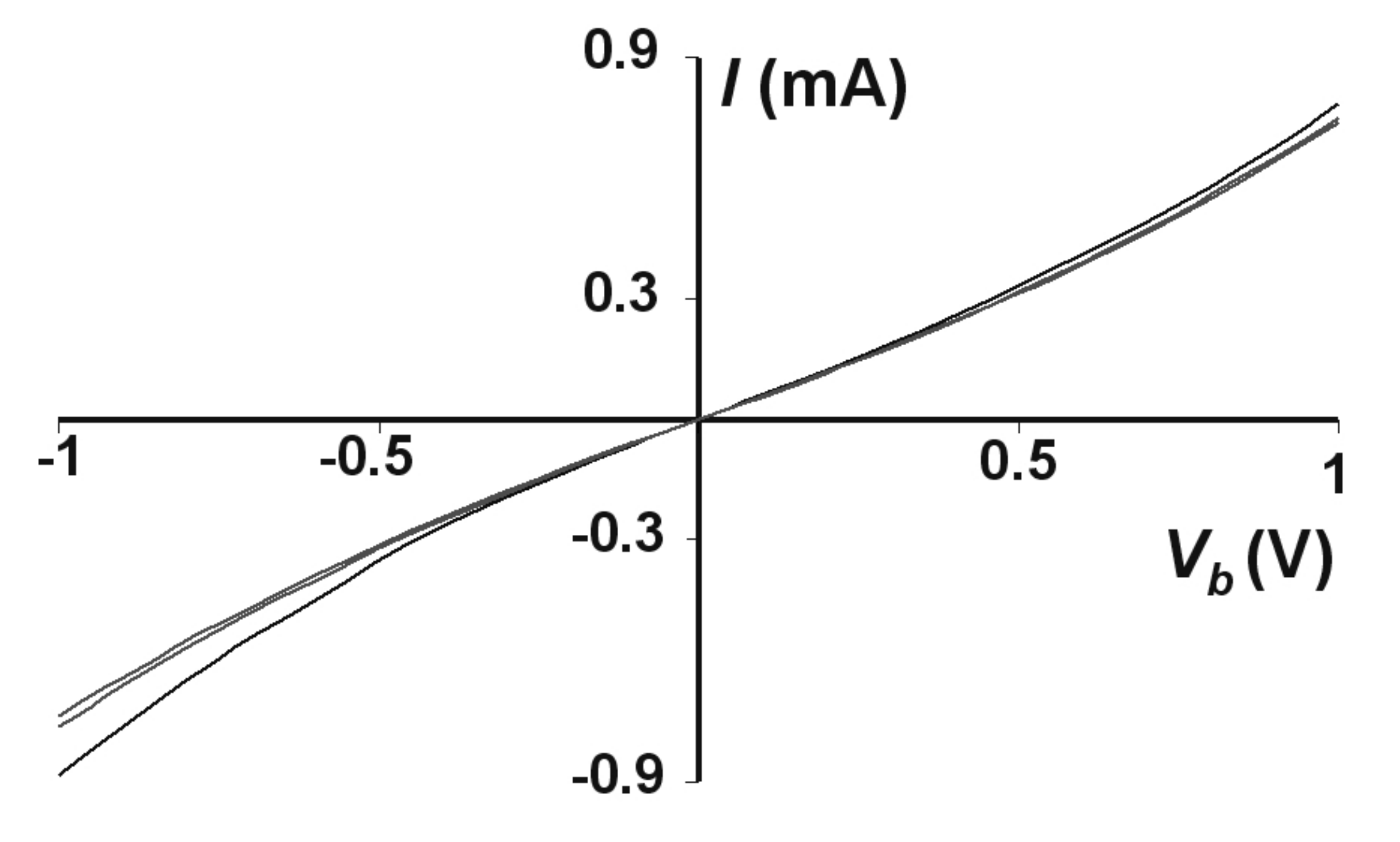}
\caption{\label{fig:epsart5} Typical Current-Voltage characteristics in forward and reverse bias of a typical acid treated CNS.}
\end{figure}
Furthermore, from the experimental curves (Fig.~\ref{fig:epsart7}) we observe for
both reference and acid treated samples a decrease of the resistance
with increasing voltage. This could be either an intrinsic effect of
the electronic structure of the CNS or could be related to local
structural modifications of the nano sheets appearing at higher bias
voltages. In fact, when we increase the voltage to about 4 V, we
observe an increase in current by one order of magnitude followed by
a drop in current close to the leakage current level of the setup which is about 50 pA. A
subsequent SEM observation demonstrates that in this case, several
10 nm large holes were created in the CNS. When the measurement was
carried out between -1 V to 1 V, no visual changes could be noted on
the CNS structures in the SEM images and, as shown above, subsequent
measurements on the same CNS show a reproducible resistivity.
Therefore, the non-linearity of the I-V curves should be an
intrinsic property of the CNS.

From the SEM images, we measured the average surface area of the CNS
in this study (Figs.~\ref{fig:epsart}(a) and ~\ref{fig:epsart}(b)). We obtain the resistivity to be 150 $\mu$ $\Omega$m
in the reference sample and 3.5 $\mu$ $\Omega$m in the acid treated
sample at 0.3V. From a comparison with resistivity values reported
in the literature, the resistivity of the reference sample
corresponds well to the value found for nanocrystalline graphite
with a crystallite size of 2-5nm.\cite{Inokawa} In fact, the CNS
studied in this work consist of structurally coherent domains or
grains with a diameter between 2-5nm in very good correspondence to
the above value.\cite{French} This similarity points to the CNS
behaving as graphite concerning their resistivity. Deviations from a
crystalline arrangement at C atoms in the grain boundaries in the
as-grown CNS will lead to an increase in the resistivity as observed
here.
\begin{figure}
\includegraphics[width=8.5cm]{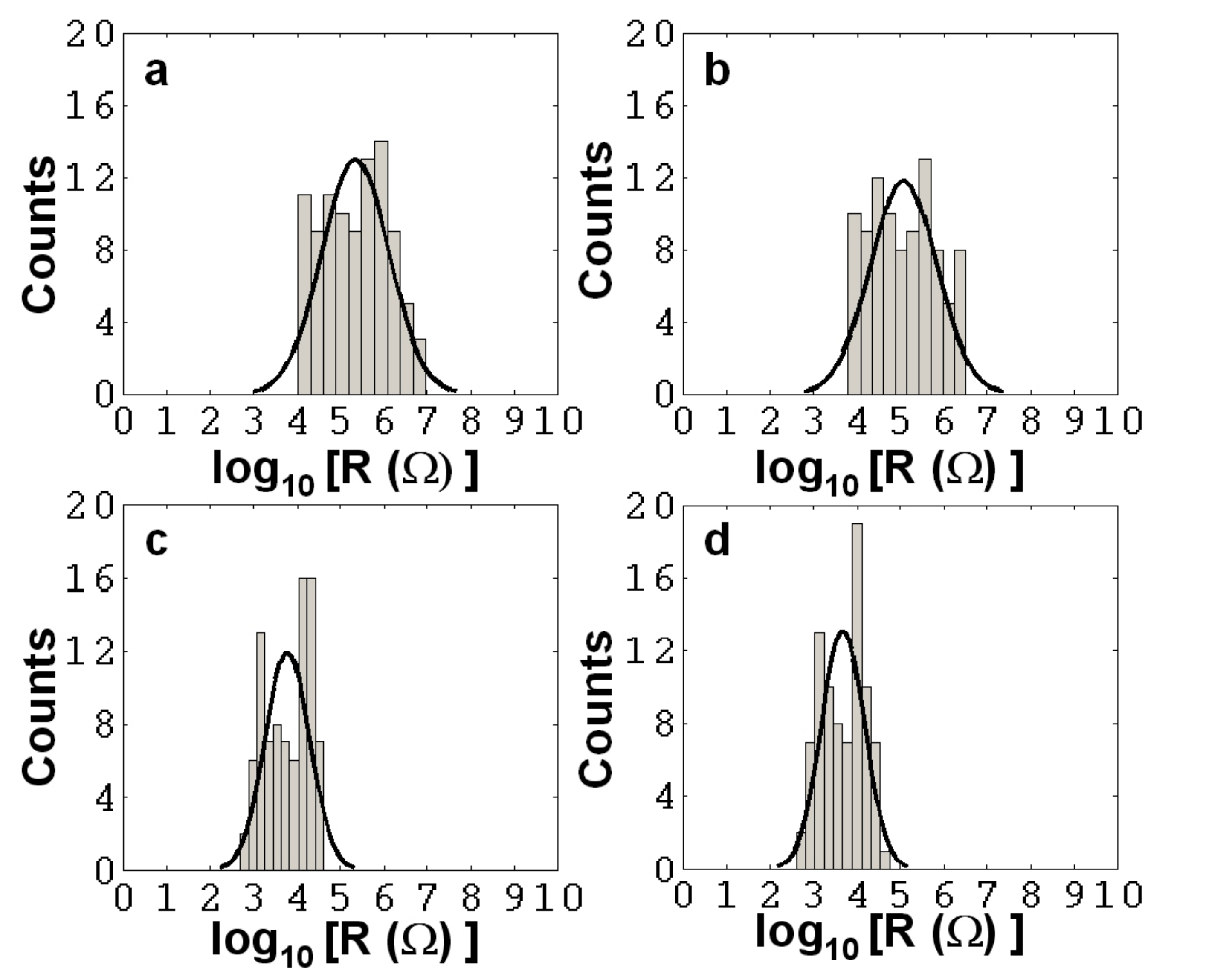}
\caption{\label{fig:epsart6} Log histograms of resistances measured in typical CNS on the reference sample at a) 0.1V b) 0.5V and on the acid treated sample at c) 0.1V d) 0.5V.}
\end{figure}
When a graphene sheet is exfoliated from a graphite crystal, the
resistivity of this sheet when positioned onto a Si substrate will
be 5 times lower, in the non-gated state, than in graphite.
\cite{bgeim} Some consequences of the acid treatment on the CNS
structure can be estimated from our observations. Structural changes
with a size of several 10 nm appear in the CNS after the acid
treatment as observed in high resolution SEM images. Our conductivity measurements show that the acid treatment has modified
the CNS structure. Such structural modification could be nucleated at
step edges present in high density (the CNS thickness changes between
2-7 monolayers) and at the grain boundaries. One may speculate
that a part of the about 50 times decrease of the resistivity upon
acid treatment could be attributed to a thinning mechanism of the
CNS, leaving on average more sheets exhibiting a graphene-like
behavior.

With this in mind, the decrease of the average resistivity from 150
$\mu$ $\Omega$m to 3.5 $\mu$ $\Omega$m measured here, is by one
order of magnitude higher than what would be expected from just a
thinning of the CNS layers and moving from a graphite character to a
graphene like conductivity. It should also be noted that the
absolute value of the non-gated average resistivity is about 3.5
$\mu$ $\Omega$m, which is as low as in measurements on low defect
graphene.\cite{bgeim} Still, the acid treated CNS will keep a
similar grain size and grain boundary defects as the reference
sample since the acid treatment is not likely to cure such defects.
Therefore, we conclude that the huge change in resistivity upon acid
treatment is largely driven by the creation of defects, and this
experimental result is consistent with the theoretical results
reported here. The presence of vacancy defects is further stressed
by our x-ray spectroscopic measurements which also show a very good
agreement with the DOS calculated for vacancy defects in graphene.\cite{coleman}
\begin{figure}
\includegraphics[width=8.5cm]{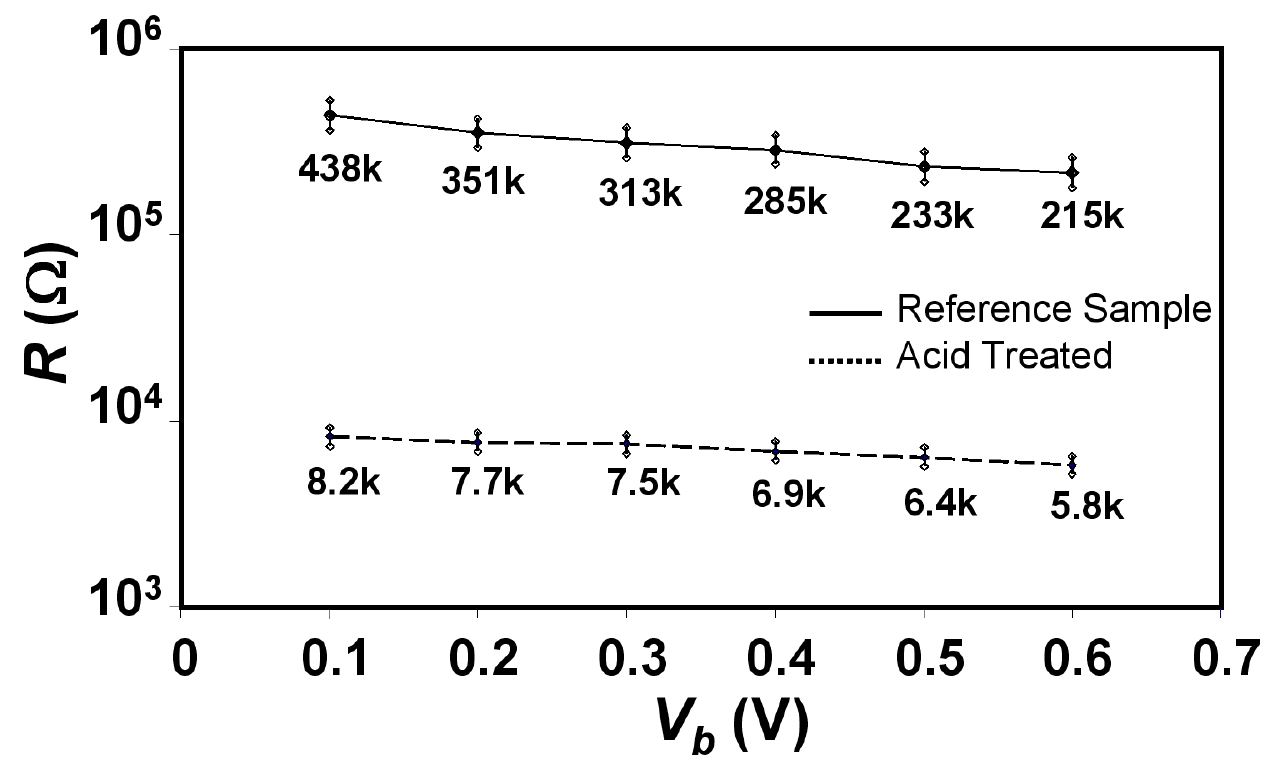}
\caption{\label{fig:epsart7}Resistance (log scale) vs. applied voltage in approx. 100
measurements (both forward and reverse bias are included) of
reference (upper curve, solid line) and acid treated sample (lower curve,
dotted line). The error bar is the standard error from the log-normal
distribution.}
\end{figure}
\section{\label{sec:level1}conclusion\protect\\}
In conclusion we have demonstrated an increase in the conductivity
of carbon nano-sheets and graphene by more than one order of
magnitude by means of engineered introduction of defects. This
behavior is drastically different from the conductivity of
defectuous carbon nano-tubes.\cite{Choi} The method to create the
defects demonstrated in this paper is very simple and widely
available, hence opening up a huge possibility of similar
investigations on graphene in various geometries, on different
substrates and in combination with other defects like adsorbed
atomic or molecular species.

\begin{acknowledgments}
The authors would like to thank A. Surpi for his
help in the set-up of the conduction measurements. We also would
like to thank the Swedish Research Council, the K. and A. Wallenberg
foundation as well as the KOF priority project at Uppsala University
and computer cluster DAVID of IOP ASCR for supporting this work.
KC acknowledges support from the Academy of Sciences of the Czech 
Republic (No. KAN400100653). BCH and RAQ acknowledge support from the U.S. AFOSR under contract
FA9550-07-C-0050. 
\end{acknowledgments}

\newpage 
\bibliographystyle{apsrev}
\bibliography{apssamp}

\end{document}